\documentclass[11pt]{article}
\usepackage{amsmath,amsfonts,latexsym,graphicx,amssymb,amsthm}
\begin{document}

\def\theequation{\thesection.\arabic{equation}}

\newcommand{\Sw}{{\cal S}(\mathbb R \frac{\ }{\ } \{ a,b \} )}

\newcommand{\mkpket}{\ensuremath{|\!-\!k^+\rangle}}
\newcommand{\pkbra}{\ensuremath{\langle ^+k|}}
\newcommand{\pebra}{\ensuremath{\langle ^+ \! \hskip0.02cm E|}}
\newcommand{\mmkbra}{\ensuremath{\langle ^- \! -k|}}

%%%%%%%%%%%%---ARROWS FOR CONVERGENCE----%%%%%%%%%%%%%%%%%%%%%%%%%%%%%%%%%%%%%
\def\llra{\relbar\joinrel\longrightarrow}              %THIS IS LONG
\def\mapright#1{\smash{\mathop{\llra}\limits_{#1}}}    %ARROW ON LINE
\def\mapup#1{\smash{\mathop{\llra}\limits^{#1}}}     %CAN PUT SOMETHING OVER IT
\def\mapupdown#1#2{\smash{\mathop{\llra}\limits^{#1}_{#2}}} %over&under it%
%%%%%%%%%%%%%%%%%%%%%%%%%%%%%%%%%%%%%%%%%%%%%%%%%%%%%%%%%%%%%%%%%%%%%%%%%%%%%%

\title{\bf Localization of non-relativistic particles}

\author{Rafael de la Madrid \\ [1ex]
\small{{\it Department of Physics, University of California at San Diego, La 
Jolla, CA 92093}} \\ [-0.5ex]
\small{E-mail: \texttt{rafa@physics.ucsd.edu}}}

\date{\small{(December 15, 2005)}}

\maketitle

\begin{abstract}
\noindent This paper is a contribution to the problem of particle 
localization in non-relativistic Quantum Mechanics. Our main results will 
be (1) to formulate the problem of
localization in terms of invariant subspaces of the Hilbert space, 
and (2) to show that the rigged Hilbert space incorporates particle
localization in a natural manner.
\end{abstract}

\vskip0.2cm

PACS numbers: 03.65.-w

\vskip0.5cm

\section{Introduction}
\setcounter{equation}{0}
\label{sec:introduction}

In Quantum Mechanics courses, we are taught that the concept of trajectory 
does not make any sense in the quantum realm. We are also taught that the 
solutions to the Schr\"odinger equation are not supposed to be interpreted 
as real waves, but rather as probability amplitudes---in Quantum Mechanics, 
what is ``waving'' is probability. We are therefore encouraged to picture 
particles not as point-like entities, but rather as sort of clouds of 
probability. This picture is reinforced by, for example, drawings of 
orbitals of the Hydrogen atom, or by animations of wave packets impinging 
upon a barrier.

We nevertheless like to think that when performing an experiment in the lab 
with, say, atoms, the wave functions that describe the atoms are localized in 
the lab. We definitely don't picture the atomic wave functions spreading all 
around space. Instead, we naively expect that we prepare clouds of 
probability that are localized in the lab, and that those clouds remain 
localized in the lab during the experiment. In this paper, we discuss to what 
extend such naive expectation holds in non-relativistic Quantum Mechanics.

Mathematically, the problem of localization can be formulated as follows.
Given a wave function $f$ that is localized at $t=0$, does $f$ remain localized
as time goes on? We shall see that such question is best formulated as the 
invariance of subspaces of the Hilbert space under the time evolution 
group. Particle localization can therefore be reduced to the study of 
invariant subspaces of the Hilbert space under time evolution.

The structure of this paper is as follows. In 
Sec.~\ref{sec:3typeslocalization}, we discuss three types of localization 
(compact-support, polynomial and
exponential) and formulate them as the invariance of certain subspaces of the
Hilbert space under time evolution. In Sec.~\ref{sec:cslocalization}, we
discuss the compact-support localization. Although it is well known that 
compact-support localization is not possible, we anyway discuss it, for the
sake of completeness and for the sake of comparison with polynomial and 
exponential localizations. In Sec.~\ref{sec:pollocalization}, we discuss the
theorems that are most relevant to polynomial localization. In 
Sec.~\ref{sec:explocalization}, we discuss exponential localization. In 
Sec.~\ref{sec:rhslocalization}, we explain how localization is built into a 
rigged Hilbert space. Finally, in Sec.~\ref{sec:conclusions}, we state our 
conclusions.

Our discussion will be elementary and, unfortunately, we shall not be able to
prove whether exponential localization holds, which is the remaining challenge
of non-relativistic particle localization.

\section{Three types of localization}
\setcounter{equation}{0}
\label{sec:3typeslocalization}

\begin{figure}[ht]
\hskip-2cm \includegraphics{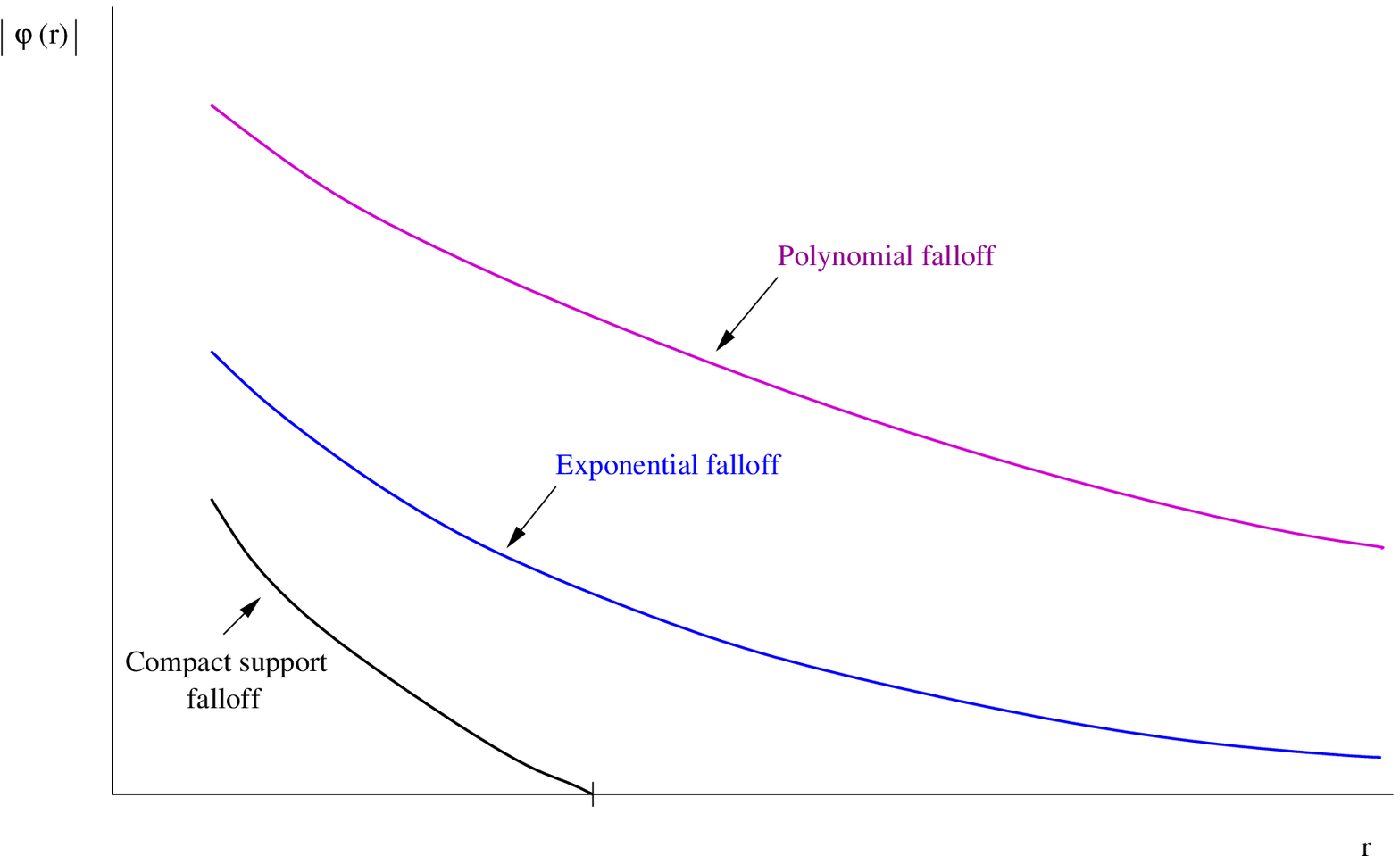}
\caption{Schematic representation of the three regimes of particle 
localization.}
\label{threetypes}
\end{figure}

Quantitatively, the localization of a particle is characterized by the rate 
at which its wave function falls off outside the region where the particle 
is supposed to be localized. There are many ways to characterize such 
falloff. The three most important falloff regimes, which are also the ones we 
are concerned with in this paper, are the compact-support, the polynomial and 
the exponential regimes (see Fig.~\ref{threetypes}):
\begin{equation}
{\rm comp. \ supp.} \prec  \cdots
\overbrace{ \prec e^{-x^n} \prec \cdots  \prec e^{-x^2} 
\prec e^{-x} \prec }^{\rm exponential \  regime}  \cdots 
\overbrace{\prec \frac{1}{x^n} \prec 
 \cdots \prec  \frac{1}{x^2}  \prec  
\frac{1}{x}}^{\rm polynomial \ regime} \, ,
\end{equation}
where $a\prec b$ indicates that the falloff ``$a$'' is stronger 
than the falloff ``$b$.'' Within each regime, one can differentiate several 
sub-regimes. For example, in the polynomial regime, one can have $1/x$ 
falloff, $1/x^2$ falloff, and so on.

In functional-analysis terms, the localization of particles can be formulated 
by constructing subspaces of the Hilbert space whose wave functions satisfy
the desired localization conditions. Thus, for compact-support localization,
we construct the space ${\mathbf \Phi}_{\rm c.s.}$ of functions $f$ that vanish
beyond a finite distance $R_f>0$:
\begin{equation}
      {\mathbf \Phi}_{\rm c.s.} = \{ f \in L^2 \, | \
        |f(x)|=0 \ {\rm for} \ |x|>R_f \, ;  \ 
           \mbox{and additional properties} \} \, , 
        \label{csphi}
\end{equation}
where in ``additional properties'' we include other extra properties that the
wave functions may have to satisfy (e.g., differentiability). For polynomial
localization of order $n$, we define the space of wave functions that
fall off faster than $x^n$:
\begin{equation}
      {\mathbf \Phi}_{\rm pol} = \{ f \in L^2 \, | \
        |x^nf(x)| \to 0 \ {\rm when} \ |x|\to \infty \, ; \ 
           \mbox{and additional properties}  \} \, . 
        \label{polphi}
\end{equation}
We can also demand polynomial localization to all orders, as with the 
Schwartz space. For exponential localization of order $n$, we
define the space of wave functions that fall off faster than
$e^{|x|^n}$:
\begin{equation}
      {\mathbf \Phi}_{\rm exp} = \{ f \in L^2 \, | \
        |e^{|x|^n}f(x)| \to 0 \ {\rm when} \ |x|\to \infty \, ;  \ 
           \mbox{and additional properties}   \} \, .
       \label{expphi}
\end{equation}
We note that the ``additional properties'' of 
Eqs.~(\ref{csphi})-(\ref{expphi}) may be necessary to ensure localization. For
example, we shall see that polynomial localization does not hold unless
additional properties are demanded from the wave functions.

Now, a particle is localized in a compact-support, polynomial or exponential
sense when the spaces ${\mathbf \Phi}_{\rm c.s.}$, ${\mathbf \Phi}_{\rm pol}$
or ${\mathbf \Phi}_{\rm exp}$ remain invariant under the time evolution 
group:
\begin{equation}
       e^{-iHt}{\mathbf \Phi}_{\rm c.s.} \subset  {\mathbf \Phi}_{\rm c.s.}
         \, , 
      \label{invariancecs}
\end{equation}
\begin{equation}
       e^{-iHt}{\mathbf \Phi}_{\rm pol} \subset  {\mathbf \Phi}_{\rm pol}
          \, , 
       \label{invariancepol}
\end{equation}
\begin{equation}
       e^{-iHt}{\mathbf \Phi}_{\rm exp} \subset  {\mathbf \Phi}_{\rm exp}
         \, .
      \label{invarianceexp}
\end{equation}
Therefore, finding out whether a particle can be localized in
a compact-support, polynomial or exponential sense is equivalent to finding
out whether the invariances~(\ref{invariancecs})-(\ref{invarianceexp}) hold 
for a given Hamiltonian. In the following three sections, we list some of the 
results that guarantee or forbid such invariances.

\section{Compact-support localization}
\setcounter{equation}{0}
\label{sec:cslocalization}

It is well known that if a non-relativistic particle is initially confined to 
a finite region of space, then it immediately develops infinite tails, as one 
could already expect from the lack of an upper limit for the propagation speed 
in non-relativistic Quantum Mechanics. Thus, compact 
support localization is impossible,
\begin{equation}
       e^{-iHt}{\mathbf \Phi}_{\rm c.s.} \subset \!\!\!\!\! /
       \  {\mathbf \Phi}_{\rm c.s.} \, .
\end{equation}

The free Hamiltonian $H_0$ provides a transparent way of seeing why
a particle initially localized in a finite region immediately spreads 
throughout all space. One simply has to calculate the time evolution of a 
wave packet $\varphi$ from the well-known expression for the free propagator
($\hbar =1$): 
\begin{equation}
      \varphi ({\bf x};t)= e^{-iH_0t}\varphi ({\bf x})= 
      \left(\frac{m}{2\pi it}\right) ^{3/2}
        \int d^3{\bf y}\  e^{im|{\bf x}-{\bf y}|^2/(2t)} \varphi ({\bf y})
       \, .
\end{equation}
The wave function $\varphi ({\bf x};t)$ is the superposition of the 
amplitudes produced by the waves emitted at $t=0$ from all points ${\bf y}$ in
space. Thus, even when $\varphi ({\bf x})$ is zero outside a finite region 
$V_0$ at $t=0$, at any other time, $\varphi ({\bf x};t)$ will be non-zero for 
all ${\bf x}$, because the free propagator ``connects'' any point ${\bf x}$
in space with those in the region $V_0$. 

An extreme case of compact-support localization occurs when the wave
function is completely supported at a point ${\bf x}_0$ of space, that is, when
the initial wave function is the delta function 
$\delta ({\bf x}-{\bf x}_0)$. For the free case, the time evolution of the 
delta function is given by
\begin{equation}
       e^{-iH_0t}\delta ({\bf x}-{\bf x}_0)= 
         \left(\frac{m}{2\pi it}\right) ^{3/2}
        \int d^3{\bf y}\  e^{im|{\bf x}-{\bf y}|^2/(2t)} 
        \delta ({\bf y}-{\bf x}_0) =
           \left(\frac{m}{2\pi it}\right) ^{3/2}
        e^{im|{\bf x}-{\bf x}_0|^2/(2t)} \, .
\end{equation}
Thus, if a free particle is initially localized at ${\bf x}_0$, then
it instantaneously develops sinusoidal tails all around space. 

\begin{figure}[ht]
\includegraphics{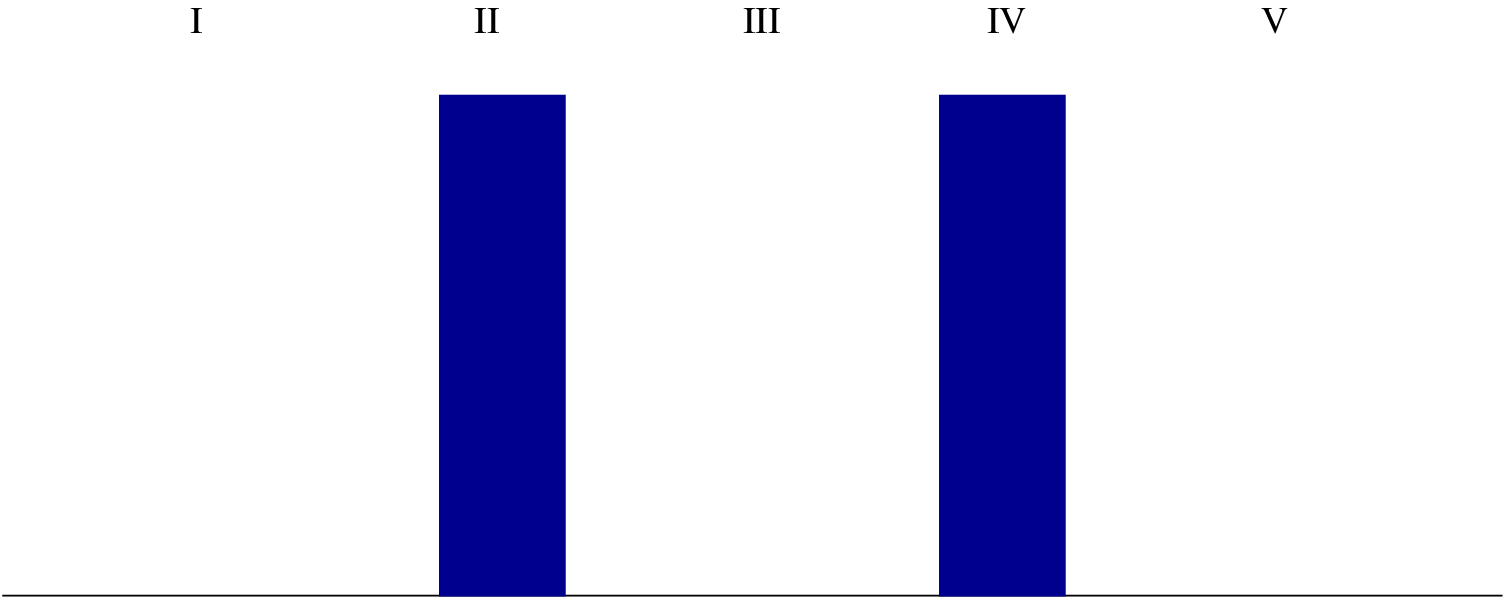}
\caption{The potential of Eq.~(\ref{potential}).}
\label{pot.eps}
\end{figure}

A theorem by Hegerfeldt~\cite{HEGERFELDT94,HEGERFELDT98} (see 
also~\cite{GALINDO68,GALINDO}) traces the impossibility of
compact-support localization to the semiboundedness of the
Hamiltonian. More precisely, if at $t=0$ the wave function is compactly
supported in a region $V_0$, and if the Hamiltonian that drives the time
evolution is bounded from below, then
\begin{enumerate}
     \item[({\it i})] either the wave function remains compactly supported 
in $V_0$,
     \item[({\it ii})] or the wave function instantaneously develops 
``tails'' that reach all regions of space. The spread is all over space, 
except for ``holes'' which, if they exist, will persist for all times.
\end{enumerate}
In most cases, possibility ({\it ii}) applies. In some cases, however,
possibility ({\it i}) applies. For example, the following potential
(see Fig.~\ref{pot.eps}) is able to trap particles in a finite
region of space:
\begin{equation}
      V(x)= \left\{ \begin{array}{ccl}
                    0 & -\infty < x < a_1  & \mbox{region I} \\
                    \infty &  a_1 < x < a_2 & \mbox{region II}  \\
                    0 & a_2 < x < a_3 & \mbox{region III}  \\
                    \infty  & a_3 < x < a_4 &  \mbox{region IV} \\
                    0 & a_4 < x < \infty  & \mbox{region V} \, . 
             \end{array} \right.  
        \label{potential}
\end{equation}
We simply have to throw the particle into region III, where it will remain
forever. This potential also illustrates the possibility of ``holes:'' If we
throw the particle in the regions I or V, then region III 
will remain as a ``hole.''

We recall that even bound states are in general not localized in a finite 
region of space. For example, the bound states of the Hydrogen atom 
fall off like an exponential multiplied by a Laguerre polynomial, and the
bound states of the Harmonic oscillator fall off like a Gaussian 
multiplied by a Hermite polynomial.

Since compact-support localization is in general not possible, the question 
now is whether the exponential and the polynomial localizations are possible, 
that is, whether ${\mathbf \Phi}_{\rm pol}$ and ${\mathbf \Phi}_{\rm exp}$ are 
invariant under the time evolution group.

\section{Polynomial localization}
\setcounter{equation}{0}
\label{sec:pollocalization}

Several theorems, especially those by Hunziker~\cite{HUNZIKER} and by 
Radin and Simon~\cite{RADIN}, guarantee 
that polynomial localization is possible when the potential is 
``reasonable.'' By ``reasonable'' we mean that there should exist an $a <1$ 
and a $b < \infty$ such that
\begin{equation}
      \| Vf \| \leq a \| H_0f \| + b \|f \|  \, ;
     \label{Katobound}
\end{equation}
that is, $V$ can be seen as a small perturbation to the kinetic energy, in
the sense of Kato~\cite{KATO}. For such potentials, we can find appropriate 
spaces ${\mathbf \Phi}_{\rm pol}$ that incorporate some type of polynomial
localization and that remain invariant under the time evolution group: 
\begin{equation}
       e^{-iHt}{\mathbf \Phi}_{\rm pol} \subset  {\mathbf \Phi}_{\rm pol}
          \, .
       \label{invariancepolp}
\end{equation}

In order to state Hunziker's theorem, we need first some definitions: 
${\bf x}^n\equiv x_1^{n_1}x_2^{n_2} x_3^{n_3}$, $n$ being the multi-index 
$({n_1},{n_2},{n_3})$ with $n_j$ integer, $n_j \geq 0$; $|n|=\sum_i n_i$; 
$k\leq n$ means $k_i\leq n_i$ for $i=1,2,3$. For any multi-index $n$, 
${\bf x}^n$ also denotes the operator multiplication by 
the function ${\bf x}^n$. For any multi-index $n$, we define a linear subset 
$D_n$ of $L^2({\mathbb R}^3)$ and a norm $\| \cdot \|_n$ on $D_n$ by
\begin{equation}
       D_n \equiv \bigcap _{\stackrel{k\leq n}{ m\leq |n|-|k|}}
             {\cal D}({\bf x}^kH^m) \, , 
        \label{Dn} 
\end{equation}
\begin{equation}
       \|f\|_n \equiv 
           \sup_{\stackrel{k\leq n}{m\leq |n|-|k|}} \|{\bf x}^kH^mf\| \, ,
       \label{normsDn}
\end{equation}
where ${\cal D}({\bf x}^kH^m)$ denotes the domain of the operator
${\bf x}^kH^m$, and $m$ denotes an integer greater than or equal to $0$.

\vskip0.3cm

{\bf Theorem~1} (Hunziker) \ {\it Under the assumption of
Eq.~(\ref{Katobound}), the following holds for any multi-index $n$:
\begin{itemize}
     \item[(a)] $D_n$ is invariant under the time evolution group:
   \begin{equation}
       e^{-iHt}D_n \subset D_n \, .
         \label{invDn}
    \end{equation}
      \item[(b)] For any $f\in D_n$, $e^{-iHt}f$ is continuous in $t$ in the 
                 sense of the norm $\| \cdot \|_n$, and there exists a 
                 constant $c_n$ such that
    \begin{equation}
       \| e^{-iHt}f\| _n \leq c_n (1+|t|)^{|n|}\| f\| _n  \, .
    \end{equation}
\end{itemize}  }

\vskip0.3cm

Since the norms of Eq.~(\ref{normsDn}) imply that the elements of $D_n$
fall off faster than $1/{\bf x}^n$ at infinity, Theorem~1 ensures the
$1/{\bf x}^n$-localization of the elements of $D_n$.

Theorem~1 is valid not only in three but in any dimension, a
result we shall take advantage of in Sec.~\ref{sec:rhslocalization}. In
addition, when the potential is a $C^{\infty}$-function with bounded
derivatives, Theorem~1 implies that the Schwartz space is invariant 
under time evolution:

\vskip0.3cm

{\bf Corollary} (Hunziker) \ {\it If $V({\bf x})$ is a bounded
$C^{\infty}$-function on ${\mathbb R}^3$ with bounded derivatives, then 
${\cal S}({\mathbb R}^3)$ is invariant under the unitary group $e^{-iHt}$ and 
the mapping $(\varphi , t)\to e^{-iHt}\varphi$ of 
${\cal S}({\mathbb R}^3)\times {\mathbb R}$ onto ${\cal S}({\mathbb R}^3)$
is continuous (in the sense of the conventional topology of
${\cal S}({\mathbb R}^3)$).}

\vskip0.3cm

Therefore, when a particle is initially localized better than any polynomial of
${\bf x}$, and when the potential that drives the evolution of the particle
is a $C^{\infty}$-function, then the particle remains localized better than 
any polynomial of ${\bf x}$ as time goes on.

A result by Radin and Simon resembles and complements Hunziker's theorem:

\vskip0.3cm

{\bf Theorem~2} (Radin-Simon) \ {\it Let $V$ obey 
Eq.~(\ref{Katobound}). Let 
\begin{equation}
       S_1 \equiv \{ f\in L^2 \, | \ |{\bf x}|f\in L^2, |P|f\in L^2\} \, ,
\end{equation}
\begin{equation}
       S_2 \equiv \{ f\in L^2 \, | \ {\bf x}^2f\in L^2, P^2f\in L^2\} \, ,
\end{equation}
and respectively equip these spaces with the norms
\begin{equation}
       \|f\|_1 \equiv \left( \|f\|^2+ \| |{\bf x}|  f\|^2 + \| |P|f\|^2 
                 \right)^{1/2} \, ,
\end{equation}
\begin{equation}
       \|f\|_2 \equiv \left( \|f\|^2+ \| |{\bf x}|^2  f\|^2 + \| P^2f\|^2
                    \right)^{1/2} \, .
\end{equation}
Then $S_1$ and $S_2$ remain invariant under $e^{-iHt}$,
\begin{equation}
       e^{-iHt}S_1 \subset S_1 \, ,
\end{equation}
\begin{equation}
       e^{-iHt}S_2 \subset S_2 \, ,
\end{equation}
and
\begin{equation}
      \| e^{-iHt}f\|_1 \leq (c+d \, |t|) \, \| f\|_1  \, ,
\end{equation}
\begin{equation}
      \| e^{-iHt}f\|_2 \leq (c'+d' t^2) \,  \| f\|_1  \, ,
\end{equation}
where $c,d,c',d'$ are constants.}

\vskip0.3cm

At infinity, the elements of $S_1$ and $S_2$ fall off faster than 
$1/|{\bf x}|$ and $1/{\bf x}^2$, respectively. Thus, Theorem~2 ensures 
the $1/|{\bf x}|$- and the $1/{\bf x}^2$-localization of the
elements of $S_1$ and $S_2$, respectively. 

Theorem~2 can be extended to higher polynomial falloffs; more 
precisely, under the conditions of Theorem~2, the space
\begin{equation}
      S_n = \{ f\in L^2 \, | \ |{\bf x}|^nf \in  L^2 \, , \ |P|^nf \in  L^2 \}
\end{equation}
is invariant under $e^{-iHt}$, for each positive $n$~\cite{OZAWA}. 

It is interesting that the falloff properties of a wave function $f$
are not preserved under $e^{-iHt}$ when $f$ has some 
singularities~\cite{OZAWA}. Thus, a wave function $f$ that is polynomially 
localized at $t=0$ will remain polynomially localized only if $f$ is smooth
enough. Hence, the space ${\mathbf \Phi}_{\rm pol}$ of
Eq.~(\ref{polphi}) always needs some ``additional properties'' in order to 
remain invariant under $e^{-iHt}$. 

There are other results on polynomial localization, all of them stating
basically that polynomial localization is possible when the wave function
is smooth enough. We shall not list all those results here; instead, we 
shall move on to the problem of exponential localization.

\section{Exponential localization}
\setcounter{equation}{0}
\label{sec:explocalization}

Contrary to polynomial localization, there doesn't seem to exist accurate
results that guarantee exponential localization of non-relativistic 
particles. Some basic results, however, indicate that exponential localization
is possible.

It is well known that a Gaussian wave packet remains Gaussian under free time
evolution. Thus, if the wave function of a free particle has Gaussian tails 
at $t=0$, and if that wave function is smooth enough, we expect that
those Gaussian tails will remain so as time goes on.

If the time evolution is driven by a Hamiltonian $H=H_0+V$, we expect that
Gaussian tails remain so as time goes on, provided that the potential $V$ is 
a small perturbation to $H_0$. 

In a scattering system, far from the potential region, the time evolution is
essentially governed by the free Hamiltonian. Thus, Gaussian tails should be 
preserved in scattering processes.

We therefore expect that for reasonable potentials and for smooth wave 
functions, exponential localization is possible. However, the precise 
statements (that is, the analogs of Theorems~1 and 2) on exponential 
localization are still lacking.

To finish this section, we note that Bialynicki-Birula has shown that the 
exponential localization of photons is possible~\cite{IBB} (see also 
Ref.~\cite{SAARI}).

\section{The rigged Hilbert space and localization}
\setcounter{equation}{0}
\label{sec:rhslocalization}

The rigged Hilbert space is emerging as the natural mathematical setting
for quantum mechanical continuous and resonance spectra. Surprisingly enough, 
the rigged Hilbert space of a system tells us a great deal about the 
localization properties of that system.

\subsection{The rigged Hilbert space and polynomial localization}

A quantum mechanical system is generally described by an algebra 
$\cal A$ of observables. These observables are defined as self-adjoint 
operators on a Hilbert space $\cal H$. More often than not, those operators 
are unbounded and have continuous spectrum, the reason for which one needs to 
construct the following rigged Hilbert spaces:
\begin{eqnarray}
      & {\mathbf \Phi}_{\rm pol} \subset {\cal H} \subset 
                     {\mathbf \Phi}_{\rm pol}^{\prime} 
                            \, , \label{RHSpolp}  \\
      & {\mathbf \Phi}_{\rm pol} \subset {\cal H} \subset 
                {\mathbf \Phi}_{\rm pol}^{\times}
                          \label{RHSpolt}  \, .
\end{eqnarray}
Here, ${\mathbf \Phi}_{\rm pol}$ is the maximal invariant subspace of the
algebra $\cal A$, and ${\mathbf \Phi}_{\rm pol}^{\prime}$ and 
${\mathbf \Phi}_{\rm pol}^{\times}$ are respectively the dual and the 
antidual spaces of ${\mathbf \Phi}_{\rm pol}$. The space 
${\mathbf \Phi}_{\rm pol}$ is the largest subspace of the Hilbert space that 
remains invariant under the action of the observables of the algebra. The 
spaces ${\mathbf \Phi}_{\rm pol}^{\prime}$ and 
${\mathbf \Phi}_{\rm pol}^{\times}$ respectively contain the bras and the
kets of the observables~\cite{EJP05,JPA04,DIS}. 

In order to see how the rigged Hilbert spaces~(\ref{RHSpolp})-(\ref{RHSpolt})
incorporate polynomial localization, we shall first consider the example of a 
spinless particle impinging on a rectangular barrier 
potential~\cite{EJP05,JPA04}. For this system, the algebra of observables is 
generated by the position, the momentum and the energy operators:
\begin{equation}
      Qf(x)=xf(x) 
       \, , \label{fdopp} 
\end{equation}
\begin{equation}
      Pf(x)=-i \frac{d}{d x}f(x) \, ,  
\end{equation}
\begin{equation}
      Hf(x)=-\frac{1}{2m}\frac{d^2}{d x^2}f(x)+V(x)f(x) \, , 
         \label{fdoph}
\end{equation}
where
\begin{equation}
           V(x)=\left\{ \begin{array}{ll}
                                0   &-\infty <x<a  \\
                                V_0 &a<x<b  \\
                                0   &b<x<\infty 
                  \end{array} 
                 \right. 
	\label{sbpotential}
\end{equation}
is the one-dimensional rectangular barrier potential. The maximal invariant
subspace of this algebra is given by a Schwartz-like space
of test functions~\cite{JPA04}, which we denote by $\Sw$. This space can be 
written as
\begin{equation}
      \Sw = \bigcap_{n=0}^{\infty}D_n \, ,
\end{equation}
with $D_n$ given by Eq.~(\ref{Dn}). The potential~(\ref{sbpotential}) 
satisfies Kato's condition~(\ref{Katobound}), because
\begin{equation}
       \|Vf\| \leq V_0 \,  \|f\| \, .
\end{equation}
We are therefore allowed to apply Theorem~1. Since by Theorem~1 each 
$D_n$ is invariant under $e^{-iHt}$, so is $\Sw$,
\begin{equation}
     e^{-iHt}  \Sw  \subset \Sw \, .
\end{equation}
This invariance, together with the polynomial falloff of the elements
of $\Sw$, ensures the polynomial localization of the elements of $\Sw$.

From the above simple example, we can draw quite general conclusions. In 
general, the algebra of a non-relativistic system will always contain the
position, the momentum and the energy operators. Hence, the elements of the 
maximal invariant subspace of the algebra, which is the space 
${\mathbf \Phi}_{\rm pol}$ of the rigged Hilbert 
spaces~(\ref{RHSpolp})-(\ref{RHSpolt}), must fall off faster than any power 
of the position coordinate. Since for a large class of 
systems Hunziker's theorem ensures the invariance of 
${\mathbf \Phi}_{\rm pol}$ under $e^{-iHt}$, the elements of 
${\mathbf \Phi}_{\rm pol}$ will in general be localized better than any 
polynomial.

It is important to note that the rigged Hilbert 
spaces~(\ref{RHSpolp})-(\ref{RHSpolt}) arise from 
properties of the algebra of the system (${\mathbf \Phi}_{\rm pol}$ is 
the maximal invariant subspace of the algebra). Therefore, the polynomial
localization built into those rigged Hilbert spaces, rather than being 
imposed by hand, arises from properties of the system.

\subsection{The rigged Hilbert space and exponential localization}

Quantum mechanical resonances are described by the Gamow states,
see e.g.~\cite{DIS,AJP02}. In the position representation, these states blow 
up exponentially at infinity. In order to control such exponential blow-up, 
we need a space ${\mathbf \Phi}_{\rm exp}$ of test functions that fall off 
faster than real exponentials~\cite{BOLLINI96,DIS}. The space 
${\mathbf \Phi}_{\rm exp}$ then yields two rigged Hilbert spaces in a natural 
way:
\begin{eqnarray}
      & {\mathbf \Phi}_{\rm exp} \subset {\cal H} \subset 
                {\mathbf \Phi}_{\rm exp}^{\prime} 
                            \, , \label{RHSexpp}  \\
      & {\mathbf \Phi}_{\rm exp} \subset {\cal H} \subset 
               {\mathbf \Phi}_{\rm exp}^{\times} \, . \label{RHSexpt}
\end{eqnarray}
Here, ${\mathbf \Phi}_{\rm exp}^{\prime}$ and 
${\mathbf \Phi}_{\rm exp}^{\times}$ are respectively
the dual and the antidual spaces of ${\mathbf \Phi}_{\rm exp}$. The space 
${\mathbf \Phi}_{\rm exp}$ is the largest subspace of the Hilbert space 
that remains invariant under the action of the observables of the algebra and 
whose elements fall off faster than any real exponential. The space
${\mathbf \Phi}_{\rm exp}^{\prime}$ contains the Gamow bras, whereas
the space ${\mathbf \Phi}_{\rm exp}^{\times}$ contains the Gamow kets.

The space ${\mathbf \Phi}_{\rm exp}$ must be invariant under $e^{-iHt}$, 
since such invariance is needed in the definition of the time evolution of the
Gamow states. Thus, the elements of ${\mathbf \Phi}_{\rm exp}$
must be exponentially localized.

It is important to realize that the Gamow states are properties of 
the Hamiltonian, and therefore so are the rigged Hilbert 
spaces~(\ref{RHSexpp})-(\ref{RHSexpt}). Hence, the exponential localization 
built into those rigged Hilbert spaces, rather than being imposed by hand, 
arises from properties of the system.

We note, however, that a satisfactory ${\mathbf \Phi}_{\rm exp}$ has not yet
been constructed for specific, simple examples. There are some proposals, 
though. For instance, Parravicini {\it et al.}~\cite{GORINI}
have proposed the space of infinitely differentiable functions with compact 
support on the positive real line, $C^{\infty}_0(0,\infty )$, as the space 
${\mathbf \Phi}_{\rm exp}$. But we saw in Sec.~\ref{sec:cslocalization} that 
$C^{\infty}_0(0,\infty )$ is not invariant under $e^{-iHt}$ for (almost) any 
$t$ and any Hamiltonian, and therefore $C^{\infty}_0(0,\infty )$ 
is inappropriate as space of test functions for the Gamow states.

\section{Conclusions}
\setcounter{equation}{0}
\label{sec:conclusions}

We have seen that the problem of localization is best formulated as the
invariance of subspaces of the Hilbert space under the time evolution 
group. We have also seen that compact-support localization is not possible,
that polynomial localization is possible, and that exponential localization
is desirable and likely to be possible. Thus, in principle, we are not able to
confine the wave packet of a particle to a finite region of space, although
we can make the tails of the wave function fall off faster than
polynomials and (probably) exponentials. 

We have also seen that the rigged Hilbert space of a system incorporates
localization in a natural way. The maximal invariant subspace of an algebra 
will in general entail polynomial localization, and the space of test 
functions for the Gamow states will in general entail exponential
localization.

So, what about our naive expectation that the wave function of our atoms
remains localized in the lab?  Do those wave functions actually spread
all around space, albeit with polynomial or exponential tails? In principle, of
course, the tails of the wave packets reach infinity. In practice, however, 
such infinity is certainly within the boundaries of the lab.

\section*{Acknowledgment}
\setcounter{equation}{0}
\label{sec:ack}

The author wishes to thank Prof.~G.~C.~Hegerfeldt for stimulating 
discussions. Correspondence with Profs.~I.~Bialynicki-Birula and
R.~de la Llave is also acknowledged. This research was supported by MEC 
fellowship No.~SD2004-0003.

\vskip3cm

\vskip1cm

\begin{thebibliography}{99}


\bibitem{HEGERFELDT94} G.~C.~Hegerfeldt, Phys.~Rev.~Lett.~{\bf 72}, 596 (1994).

\bibitem{HEGERFELDT98} G.~C.~Hegerfeldt, ``Causality, Particle Localization
and Positivity of the Energy,'' in ``Irreversibility and Causality,'' 
A.~Bohm, H.-D.~Doebner, P.~Kielanowski [Eds.], Springer Lecture Notes in
Physics, Springer (1998).

\bibitem{GALINDO68} A.~Galindo, Anales de F\'\i sica~{\bf 64}, 141 (1968).

\bibitem{GALINDO} A.~Galindo, P.~Pascual, \emph{Mec\'anica Cu\'antica},
Universidad-Manuales, Eudema (1989); English translation by J.~D.~Garc\'\i a
and L.~Alvarez-Gaum\'e, Springer-Verlag (1990).

\bibitem{HUNZIKER} W.~Hunziker, J.~Math.~Phys.~{\bf 7}, 300 (1966).

\bibitem{RADIN} C.~Radin, B.~Simon, J.~Differential Equations~{\bf 29}, 289
(1978).

\bibitem{KATO} T.~Kato, \emph{Perturbation Theory for Linear 
Operators}, Springer Verlag, New York (1966).

\bibitem{OZAWA} T.~Ozawa, Archive for Rational Mechanics and 
Analysis~{\bf 110}, 165 (1990).

\bibitem{IBB} I.~Bialynicki-Birula, Phys.~Rev.~Lett.~{\bf 80}, 5247 (1998).

\bibitem{SAARI} P.~Saari, M.~Menert, H.~Valtna, ``Photon localization barrier
can be overcome,'' {\sf quant-ph/0409034}.

\bibitem{EJP05} R.~de la Madrid, Eur.~J.~Phys.~{\bf 26}, 287 (2005);
{\sf quant-ph/0502053}. 

\bibitem{JPA04} R.~de la Madrid, J.~Phys.~A:~Math.~Gen.~{\bf 37}, 8129 (2004);
{\sf quant-ph/0407195}. 

\bibitem{DIS} R.~de la Madrid, {\it Quantum Mechanics in Rigged
Hilbert Space Language}, Ph.D.\ thesis, Universidad de Valladolid,
Valladolid, 2001. Available at 
\texttt{http://www.physics.ucsd.edu/$\sim$rafa/}.

\bibitem{AJP02} R.~de la Madrid, M.~Gadella, Am.~J.~Phys.~{\bf 70}, 626 (2002);
{\sf quant-ph/0201091}. 

\bibitem{BOLLINI96} C.~G.~Bollini, O.~Civitarese, A.~L.~De Paoli, M.~C.~Rocca,
Phys.~Lett.~B{\bf 382}, 205 (1996); J.~Math.~Phys.~{\bf 37}, 4235 (1996).

\bibitem{GORINI} G.~Parravicini, V.~Gorini, E.~C.~G.~Sudarshan,
J.~Math.~Phys.~{\bf 21}, 2208 (1980).


\end{thebibliography}
\end{document}